\documentclass{ws-mpla}

\newcommand{\bea}{\begin{eqnarray}}
\newcommand{\eea}{\end{eqnarray}}
\newcommand{\be}{\begin{equation}}
\newcommand{\ee}{\end{equation}}

\newcommand{\rt}[1]{{}}
\newlength{\szovszel}
\newlength{\slashszel} 
\newcommand*{\sls}[1]{\mbox{%
   \settowidth{\szovszel}{\ensuremath{#1}}%
    \settowidth{\slashszel}{\ensuremath{\slash}}%
    \hspace*{0.5\szovszel}%
    \hspace*{-0.5\slashszel}%
    \slash%
    \hspace*{-0.5\szovszel}%
    \hspace*{-0.5\slashszel}%
    \ensuremath{#1}%
  }}

\begin{document}

\markboth{A. Jakov\'ac I. Kaposv\'ari A. Patk\'os}
{Scalar mass stability bound }

%%%%%%%%%%%%%%%%%%%%% Publisher's Area please ignore %%%%%%%%%%%%%%
\catchline{}{}{}{}{}
%%%%%%%%%%%%%%%%%%%%%%%%%%%%%%%%%%%%%%%%%%%%%%%%%%%%%%%%%%%%%%%%%%%

\title{Scalar mass stability bound in a simple Yukawa-theory from
  renormalisation group equations 
}

\author{\footnotesize A. Jakov\'ac}

\address{Department of Atomic Physics, E\"otv\"os University \\
H-1117 P\'azm\'any P. s\'et\'any 1/A\\ Budapest, Hungary\\jakovac@caesar.elte.hu
}

\author{\footnotesize I. Kaposv\'ari}
\address{E\"otv\"os Univerity, H-1117 P\'azm\'any P. s\'et\'any 1/A\\ Budapest, Hungary\\
stevenkaposvari@gmail.com}

\author{A. Patk\'os}

\address{MTA-ELTE Research Group for Biological and Statistical Physics\\
H-1117 P\'azm\'any P. s\'et\'any 1/A\\ Budapest, Hungary\\patkos@galaxy.elte.hu
}

\maketitle

\pub{Received (Day Month Year)}{Revised (Day Month Year)}

\begin{abstract}
Functional Renormalisation Group (FRG) equations are constructed for a
  simple Yukawa-model with discrete chiral symmetry,
  including also the effect of a nonzero composite fermion background
  beyond the conventional scalar condensate. The evolution of the effective potential of the model, generically depending on two invariants, is explored with help of power series expansions. Systematic investigation of the effect of a class of irrelevant operators on the lower (stability) bound allows a non-perturbative extension of the maximal cut-off value consistent with any given mass of the scalar field.

\end{abstract}

\ccode{PACS Nos.:11.10.Gh; 11.10.Hi}

\section{Introduction}	
The experimentally found mass of the Higgs-field is light in the sense that it is very close to the lower edge of the perturbatively determined infrared window (for recent theoretical reviews, see \cite{espinosa13,eichhorn15}). This situation resulted in an increased number of studies investigating the influence of quantum fluctuations on the stability of the Higgs effective potential. 
For the characterisation of this influence two conceptually different interpretations were put forward. The first is motivated by the setup of Monte Carlo simulations of the ground state of field theories where one finds no signal for any instability in the continuum limit if the theory defined at the scale of the lattice constant (approximately the inverse momentum cutoff) is stable \cite{holland04,holland05,fodor07,gerhold09,gerhold10}. Along this line of thought one arrives for any fixed value of the scalar mass at a maximal value of the cut-off beyond which no stable bare Lagrangian can exist with a fixed set of operator content.

The alternative interpretation restricts the bare potential energy of the scalar field to the perturbatively renormalisable quartic power. One investigates the renormalisation group equation (RGE) of the renormalized quartic coupling embedded into the standard model as a function of the renormalisation scale when the cut-off is sent to infinity \cite{krive76,linde80,lindner89,degrassi12,butazzo13,branchina15}. One encounters a negative value for this coupling upon fixing its "measured" value at some infrared scale, when the renormalisation scale goes beyond a certain, maximally allowed value. The order of magnitude of this "critical" value is the same as of the maximal cutoff value emerging from the first approach. 

The close agreement of the stability (lower) mass bounds obtained using the set of perturbatively renormalisable operators with various techniques ranging from perturbation theory, through semianalytic solution of the functional renormalisation group equations to the lattice simulations give considerable mutual support to the credibility of the results. 

Recently adding perturbatively irrelevant operators to the Higgs potential was shown to increase the maximal allowed cutoff considerably\cite{gies14,gies15,chu15}. It is not clear yet if there exist any absolute lower bound in this wider sense, since no systematic exploration of the coupling space has been conducted to date. 

In the present paper we extend the operator set involved in the determination of the lower mass bound with the FRG. Earlier approximate computations were restricted to a one-variable effective potential depending on the invariant $\rho$ built exclusively from the scalar field. 
We shall consider the RG-evolution of the potential energy
depending also on a second variable, the local invariant
formed jointly with the scalar field and the fermions. The effect of this new invariant (called below the Yukawa-invariant, and denoted by $I$) will be investigated in the framework  of the simplest scalar-fermion Yukawa theory \cite{gies14}. 

Actually, we shall study the linear dependence on the Yukawa-invariant. Such an extension of the scalar-fermion action has been considered first in Ref.\refcite{pawlowski14} in the context of a quark-meson theory, where it emerges naturally from the RG-flow. For strong enough scalar self-coupling similar dynamical generation might manifest itself in the present system. In first place, the point where the two-variable potential is expanded to linear order in the Yukawa-invariant will be varied. Apparently, this freedom leads to an almost negligible variation in the scalar
mass bound at fixed cutoff. This feature justifies the omission of operators involving higher powers of $I$ from the ansatz. The evolution of all couplings of the quantum action follows rather closely the rate determined by the perturbatively computed respective beta-functions until only perturbatively relevant and marginal operators are considered. The lower bound characteristic for this case is reached when the initial value of the quartic scalar self-coupling is chosen zero. 

The situation changes spectacularly when one allows the coefficient in front of the Yukawa-invariant (the Yukawa-coupling) to depend on the scalar field invariant, introducing this way perturbatively irrelevant operators into the potential. In this case the evolution of the couplings determined by the RG-equations joins with a considerable delay the nearly perturbative regime after a non-perturbative "prehistory". Initial data for the extra evolution are further optimised dynamically by running towards the ultraviolet from the optimal initial data found when restricting the initial quartic scalar coupling to zero. This running is stopped now when the model looses its stability through the generalised Yukawa coupling becoming negative. This procedure is used to map the curve of the lowest mass estimate found for field independent Yukawa coupling onto a new curve. This way one extends the range of allowed cutoff momenta to values larger at least by one order of magnitude.

The paper is organised as follows. In section 2 the model and the Ansatz for  its scale dependent effective action are introduced. The set of RGE's is constructed explicitly in presence of scalar and fermionic backgrounds in the formulation due to Wetterich \cite{wetterich91,wetterich93} (see also Ref.\refcite{morris94}). The extra contribution reflecting the presence of the fermionic background is clearly identified. In section 3 the RGE's are constructed for the effective potential expanded in the Yukawa-invariant to linear order at an arbitrary point. Here we also describe the strategy for scalar mass estimates from the lowest non-trivial power-function ansatz for the potential. In section 4 the numerical solution will be presented in the Local Potential Approximation (LPA) of the Wetterich equation. We shall concentrate on the effects arising from the choice of the expansion point and from including the specific, perturbatively irrelevant piece into the Yukawa-coupling depending linearly on the scalar invariant. In section 5 the robustness of the effect of this operator is thoroughly investigated when one steps beyond LPA taking into account the anomalous dimensions of the fields (${\textrm{LPA}}^\prime$) and also further higher powers are included into the field dependence of the Yukawa coupling and in the scalar potential. A summary and the conclusions of this investigation appear in Section 6.

\section{The model and the application of the Renormalisation Group}

The scalar-fermion Yukawa model in Euclidean space-time will be treated in the framework of the following effective action Ansatz defining the theory at scale $k$:
\begin{equation}
\Gamma_k=\int_x\left[Z_{\psi k}\bar\psi\gamma_m\partial_m\psi+
\frac{1}{2}Z_{\sigma k}(\partial_m\sigma)^2+W_k(\rho,I)\right],\qquad \rho=\frac{1}{2}\sigma^2,\quad I=\sigma\bar\psi\psi.
\label{action-ansatz}
\end{equation}
Here the last term represents the potential energy density of the model, which we shall expand to linear order in $I$: $W_k\approx U_k(\rho)+h_k(\rho)I$. A new feature of the present investigation relative to Ref.\refcite{gies14} is the general $\rho$-dependence of the Yukawa-type coupling. The model possesses a discrete $\gamma_5$ invariance:
\begin{equation}
\sigma(x)\rightarrow -\sigma(x),\qquad \psi(x)\rightarrow \gamma_5\psi(x),\quad \bar\psi(x) \rightarrow -\bar\psi(x)\gamma_5.
\label{gamma5-symmetry}
\end{equation}

A non-zero homogeneous condensate at scale $k$ given as $v_k=Z_{\sigma k}^{1/2}\sigma_k=\sqrt{2Z_{\sigma k}\rho_k}$ in the scalar field $\sigma$ corresponds to the spontaneous symmetry breaking of the discrete $\gamma_5$ transformation. 
The condensate generates mass for the fermi-field:
$m_{\psi,k}=h_k(\rho_k)\sigma_k$, which itself varies with the scale. The
 physical condensate and the
physical mass values are computed from the action $\Gamma_{k=0}$, at
which the starting $\Gamma_{k=\Lambda}$ will land after running from
the cutoff $\Lambda$ to $k=0$:
\begin{eqnarray}
&
v=Z_{\sigma 0}^{1/2}\sigma_0,\qquad m_\psi^2=2h_0^2(\rho_0)\rho_0,\nonumber\\
&
 m_\sigma^2=U'_0(\rho_0)+2\rho_0U''_0(\rho_0)+(3h_0'(\rho_0)+2\rho_0h_0''(\rho_0))I_0\equiv U_0^{(1)}(\rho_0)+h_0^{(2)}(\rho)I_0.
\label{mass-condensate}
\end{eqnarray}
To ensure maximal analogy between our model and the top-Higgs sector of the Standard Model, in the discussion of the solution, below we fix the $k=0$ values of the vacuum condensate to $v=246$ GeV and the fermion mass to $m_\psi=173$ GeV. 

The RGE\cite{wetterich91,wetterich93} as projected on (\ref{action-ansatz})
expresses the rate of variation of $\Gamma_k$ as a function of $t=\ln(k/\Lambda)$ through an expression containing $\Gamma^{(2)}_k$, the second functional derivative matrix of the effective action. It is represented as a 3x3 matrix,
where the fields are represented by the transposed of the row-vector   $(\sigma(x), \psi^T(x), \bar\psi(x))$ introducing the Gor'kov-Nambu representation of the fermi-field. One can transform  this matrix into a block-diagonal form \cite{jakovac13}, which leads for the second equality of RGE:
\begin{eqnarray}
&&\partial_t\Gamma_k[\Phi]=\frac{1}{2}\hat\partial_t{\textrm{Str}}\log(\Gamma^{(2)}+R_k)\nonumber\\
&&=
-\frac{1}{2}\hat\partial_t{\textrm{Tr}}\log(\Gamma^{(2)}_{\Psi^T\Psi}+R_k^F)+\frac{1}{2}\hat\partial_t{\textrm{Tr}}\log(\Gamma^{(2)}_{\sigma\sigma}+R_k^B)\nonumber\\
&&
+\frac{1}{2}\hat\partial_t{\textrm{Tr}}\log\Biggl[1-\frac{1}{\Gamma^{(2)}_{\sigma\sigma}+R_k^B}\biggl[\Gamma^{(2)}_{\sigma\psi}\frac{1}{\Gamma^{(2)}_{\bar\psi\psi}+R_k^F}\Gamma^{(2)}_{\bar\psi\sigma}+\Gamma^{(2)}_{\sigma\bar\psi^T}\frac{1}{\Gamma^{(2)}_{\psi^T\bar\psi^T}+R_k^F}\Gamma^{(2)}_{\psi^T\sigma}\biggr]\Biggr].
\label{super-tracelog-1}
\end{eqnarray}
$R_k$ is the infrared regulator restricting the contribution to the functional trace $\textrm{Str}$ onto the neighbourhood of the actual momentum scale $k$. The symbol $\hat\partial_t$ refers to a $t$-derivative acting exclusively on $R_k$. Throughout this paper we shall employ the so-called optimised regulator \cite{litim01}. The symbol $\Gamma^{(2)}_{\Psi^T\Psi}$ denotes a hypermatrix in the doubled bispinor field basis  where the condensed notation $\Psi^T=(\psi^T,\bar\psi)$ and its transposed column-vector is used. 
\begin{figure}
\begin{center} 
\includegraphics[width=3cm]{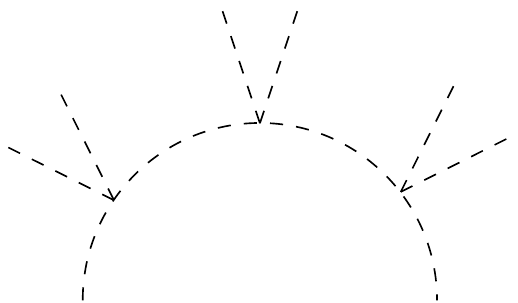}
\hspace*{1cm}
\includegraphics[width=3cm]{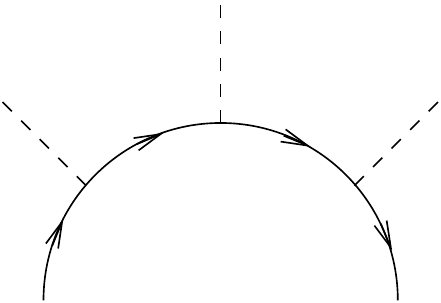}
\hspace*{1cm}
\includegraphics[width=3cm]{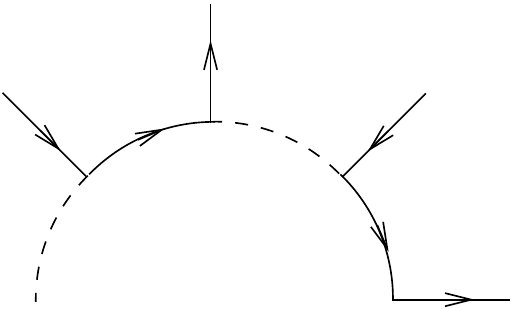}
\end{center}
\caption{Propagator sequences building the one-loop contributions to the tracelogs of (\ref{super-tracelog-1}). On the left a piece of a scalar loop (dashed curve), in the middle a fermion loop (continuous curve) appear both immersed into a scalar condensate, on the right a loop constructed of alternating sequence of fermion and scalar propagators immersed into a composite fermionic condensate is represented.}
\label{one-loop-contribution-tracelog}
\end{figure}
Each term has a diagrammatic representation (see Fig.\ref{one-loop-contribution-tracelog}). The first two terms correspond to the free fermion and free boson contributions to the super-tracelog computed in the background $\sigma_k$. The last term is the sum of the infinite series of 1-loop contributions containing an increasing number of alternating boson and fermion propagators. At each vertex of fermion-to-boson transformation an external fermion leg joins the loop, reflecting the composite fermionic background . 

In earlier papers the super-trace was evaluated on a background exclusively consisting of a nonzero $\sigma$-condensate \cite{gies14}, where $\Gamma^{(2)}_{\sigma\Psi}=\Gamma^{(2)}_{\Psi^T\sigma}=0$.
In our recent publications clear arguments were put forward \cite{jakovac13,jakovac15} how the Grassmannian nature of the fermi fields is compatible with a coarse grained background where also nonzero
$[(\bar\psi\psi)_k]^n, ~n>1$ terms are present. Earlier this possibility was studied in a series of papers also by Aoki {\it et al.} \cite{aoki00a,aoki00b,aoki13,aoki14} . 

\section{Renormalisation Group Equations in Local Potential Approximation (LPA)}

\subsection{The potential energy}

For this calculation one chooses from the start space-time independent scalar and fermi backgrounds. On this background one has the following representations for the propagators using linear infrared regulator functions $r_B(k/p), r_F(k/p)$ \cite{litim01}:
\begin{equation}
G^{-1}_\psi(p)=iZ_\psi P_F(p)\sls p-m_{\psi,k},\quad P_F=1+r_F,\quad r_F(p)=\left(\frac{k}{\sqrt{p^2}}-1\right)\Theta\left(\frac{k^2}{p^2}-1\right) ,
\end{equation}
and
\be
G_\sigma^{-1}(p)=Z_\sigma P_B(p)p^2+m_{\sigma,k}^2,\quad P_B=1+r_B,\quad r_B(p)=\left(\frac{k^2}{p^2}-1\right)\Theta\left(\frac{k^2}{p^2}-1\right).
\ee
 The masses are given with the expressions in (\ref{mass-condensate}), but computed with the corresponding background fields and couplings defined at scale $k$.
 
Below the notation $q_R^2$ will be used as a short-hand symbol for the momenta appearing in combinations $Z_\sigma P_B(q)q^2$ or $Z_\psi^2 P_F^2(q)q^2$, depending on whether one deals with the bosonic or fermionic propagator. One finds with a few algebraic steps the following expression for the rate of evolution of the potential energy density ($h^{(1)}_k(\rho_k)\equiv h_k(\rho_k)+2\rho_k h'(\rho_k)$):
\bea
&&\partial_t[U_k(\rho_k)+h_k(\rho_k)I_k)]=\nonumber\\
&&\frac{1}{2}\hat\partial_t\int_q\left[\log\frac{q_R^2+m_\sigma^2}{(q_R^2+m_\psi^2)^4}+\log\left\{1-\frac{h_k(\rho)(h_k^{(1)}(\rho_k))^2}{q_R^2+m_\sigma^2}\frac{2I_k}{q_R^2+m_\psi^2}\right\}\right].
\label{LPA-RGE}
\eea
In view of the general $I_k$-dependence of the right hand side, present also through $m_{\sigma,k}^2$, the right hand side is to be projected on the ansatz linear in $I_k$ figuring on the left hand side. In fact, one can use an arbitrary function $I_k^s(\rho_k)$ as expansion point: 
\be
{\textrm {RHS}}(\rho_k,I_k)={\textrm {RHS}}(\rho_k,I^s_k(\rho_k))+(I_k-I^s_k(\rho_k))\frac{\partial}{\partial I_k}{\textrm {RHS}}(\rho_k,I^s_k(\rho_k))+...,
\label{RHS-expansion}
\ee
The projection has been realised in earlier investigations by an expansion around the origin $I_k^s=0$ \cite{gies14}.
If one would know the full two-variable potential $W(\rho,I)$ then a natural choice would be $I_k^s=I_{min}(\rho_k)$ determined by the condition $\partial W/\partial I=0$. 
Such a procedure is analogous to the fermion-boson translation strategy of Ref.\refcite{gies02}. 

Keeping both
terms displayed on the right hand side of (\ref{RHS-expansion}) and writing separately the RGE for the expression linear in $I_k$ one has
\bea
\partial_tU_k(\rho_k)
&=&\frac{1}{2}\hat\partial_t\int_q\Bigl\{-5\log(q_R^2+2h_k^2\rho_k)\nonumber\\
&&+\log\left[(q_R^2+U_k^{(1)}+h_k^{(2)}I^s_k(\rho_k))(q_R^2+2h_k^2\rho_k)-2h_k(h_k^{(1)})^2I^s_k(\rho_k)\right]\Bigr\}\nonumber\\
&&
+\frac{1}{2}\hat\partial_t\int_q\frac{2h_k(h_k^{(1)})^2-h_k^{(2)}(q_R^2+2h_k^2\rho_k)}{(q_R^2+U_k^{(1)}+h_k^{(2)}I^s_k(\rho_k))(q_R^2+2h_k^2\rho_k)-2h_k(h_k^{(1)})^2I_k^s}I_k^s,
\label{LPA-version-2a}
\eea
\be
\partial_t(h_k(\rho_k)I_k)=-\frac{1}{2}\hat\partial_t\int_q\frac{2h_k(h_k^{(1)})^2 -h_k^{(2)}(q_R^2+2h_k^2\rho_k)}{(q_R^2+U_k^{(1)}+h_k^{(2)}I^s_k(\rho_k))(q_R^2+2\rho_kh_k^2)-2h_k(h_k^{(1)})^2I_k^s}I_k.
\label{LPA-version-2b}
\ee 
When one chooses $\rho$-independent Yukawa coupling and $I_k^s=0$, the equations coincide with the set of RGE's of derived in Ref.\refcite{gies14}. Still, the RGE of $h_k(\rho)$ alone is different depending on whether one defines it as $\Gamma^{(3)}_{\bar\psi\psi\sigma}$ \cite{gies14} or as $\partial W(\rho,I)/\partial I$ as we do in the present study (see below).
In both cases when one constructs a power series solution for $h_k(\rho_k)$, then the most convenient is to expand it around $\rho_{k,min}$, since the physical fermion mass is determined with $h_0(\rho_{0,min})$. 

For the solution of the RGE's, they are rewritten in terms of dimensionless background field variables and couplings, rescaled by appropriate powers of the scale $k$. These quantities are distinguished by the lower index 'r'.
The operation $k\hat\partial_k$ is performed for the linear regulator\cite{litim01}.
 The coupled equations of the potential and the Yukawa-coupling are as follows, after cancellation of $I_r$ on both sides of the latter:
\bea
&&
\partial_tU_r(\rho_r)+dU_r(\rho_r)+(2-d-\eta_\sigma)\rho_rU'_r(\rho_r)\nonumber\\
&&
=v_d\left[-\frac{5M_\psi}{1+\mu_\psi^2}+\frac{(1+\mu_\sigma^2)M_\psi+(1+\mu_\psi^2)M_\sigma}{(1+\mu_\psi^2)(1+\mu_\sigma^2)-2h_r(h_r^{(1)})^2I_r(\rho_r)}\right]\nonumber\\
&&+v_d\frac{I^s_r(\rho_r)}{[(1+\mu_\psi^2)(1+\mu_\sigma^2)-2h_r(h_r^{(1)})^2I_r^s]^2}\times\nonumber\\
&&
\left[(h_r^{(2)}(1+\mu_\psi^2)-2h_r(h_r^{(1)})^2)
(1+\mu_\psi^2)M_\sigma
%\nonumber\\
%&&~~~~~~~~~~~~~~~~~~~
-2h_r(h_r^{(1)})^2(1+\mu_\sigma^2-h_r^{(2)}I_r^s)M_\psi\right],
\label{Z-RGE-B1}
\eea
\bea
&&\partial_th_r+(2-d-\eta_\sigma)\rho_rh_r'+\frac{1}{2}(4-d-\eta_\sigma-2\eta_\psi)h_r
\nonumber\\
&&=-\frac{v_d}{2[(1+\mu_\psi^2)(1+\mu_\sigma^2)-2h_r(h_r^{(1)})^2I_r^s]^2}\times\nonumber\\
&&
\left[(h_r^{(2)}(1+\mu_\psi^2)-2h_r(h_r^{(1)})^2)
(1+\mu_\psi^2)M_\sigma
-2h_r(h_r^{(1)})^2(1+\mu_\sigma^2-h_r^{(2)}I_r^s)M_\psi\right].
\label{Z-RGE-B2}
\eea
Here the notations $v_d=S_d/(d(2\pi)^d),~~v_4=(32\pi^2)^{-1}$, $\eta_\sigma=-\partial_t\ln Z_\sigma, \eta_\psi=-\partial_t\ln Z_\psi$  and $M_\sigma=1-\eta_\sigma/(d+2), M_\psi=1-\eta_\psi/(d+1)$ are introduced, with $S_d$ standing for the surface of the $d$-dimensional unit sphere. The boson 
 (fermion) masses are rescaled by $Z_\sigma k^2~  (Z_\psi^2k^2)$ to give  $\mu_\sigma^2~~ (\mu_\psi^2)$.

In order the reader could appreciate the non-trivial numerical uniformity of the lower bounds presented below, the extra terms which would show up on the RHS of (\ref{Z-RGE-B2}) when one defines $h_r$  as $\Gamma^{(3)}_{\bar\psi\psi\sigma}$  are also given:
\bea
&&+\hat\partial_t\int_q\left[\frac{\rho h (h^{(1)})^2(3U''+2\rho U''')}{(q_R^2+U^{(1)})^2(q_R^2+2h^2\rho)}+\frac{2h^2(h^{(1)})^3}{(q_R^2+U^{(1)})(q_R^2+2h^2\rho)^2}\right]\nonumber\\
&&
-\hat\partial_t\int_q\left[\frac{1}{2}\frac{(h^{(1)})^3+4\rho hh^{(1)}h^{(2)}}
{(q_R^2+U^{(1)})(q_R^2+2h^2\rho)}-\frac{\rho (h^{(2)})'}{q_R^2+U^{(1)}}+\frac{\rho h^{(2)}U^{(1)'}}{(q_R^2+U^{(1)})^2}\right].
\eea

\subsection{Power series analysis for $U_r$ and $h_r$}

The simplest search for the solutions makes use of power series of $U_r$ and $h_r$
and sets $Z_\psi=Z_\sigma=1$.  One finds nontrivial results already with a quadratic expression for $U_r$ which corresponds to keeping in the effective action only perturbatively relevant+marginal operators. The flexibility of the lower bound under some modulation of the Yukawa coupling will be investigated with an ansatz linear in $\rho_r$: $h_r=h_0+h_1\rho_r$. In the symmetric phase one has 
\be
U_r(\rho_r)=\lambda_{1k}\rho_r+\frac{1}{2}\lambda_{2k}\rho_r^2,\quad \mu_\psi^2=2(h_0+h_1\rho_r)^2\rho_r,\quad
\mu_\sigma^2=\lambda_{1k}+3\lambda_{2k}\rho_r+3h_1I_r^s,
\label{pot-SYM}
\ee
($ \lambda_{1k}>0$), while the solution in the broken symmetry phase is parametrized as
\bea
&\displaystyle
U_r(\Delta\rho=\rho_r-\rho_{r,min})=\frac{\lambda_{2k}}{2}(\Delta\rho)^2,
\qquad \mu_\sigma^2=2\lambda_{2k}\rho_{r,min}+3\lambda_{2k}\Delta\rho+3h_1I_r^s,\nonumber\\
&\displaystyle
\mu_\psi^2=2\bar h^2\rho_{r,min}+2(\bar h^2+ 2\bar h h_1\rho_{r,min})\Delta\rho+
2h_1(h_1\rho_{r,min}+2\bar h)\Delta\rho^2+2h_1^2\Delta\rho^3,
\label{pot-SBSYM}
\eea
with $\rho_{r,min}$, the rescaled minimum of $U_r$, ($U'_r(\rho_{r,min})=0$) and $\bar h=h_0+h_1\rho_{r,min}$.

The above mentioned two cases for the choice of $I_r^s(\rho_r)$ were compared.
 After expanding the right hand side of RGE in power series of $\rho_r$ or $\Delta\rho$ and equating the corresponding terms one arrives at a set of equations for the couplings $\lambda_{nk}$ and/or the expectation value $\rho_{r,min}$. The consistency of the two sides are ensured by keeping terms on the RHS of 
  (\ref{Z-RGE-B1}) to quadratic, while on the RHS of (\ref{Z-RGE-B2}) to first order in $\rho_r$ or $\Delta\rho$.

The discussion of the influence of the truncation at higher (perturbatively irrelevant) couplings of $U_r$ on
the stability bound requires some care since truncations at odd powers lead to seemingly unstable potentials with the coefficients in front of the highest power converging to zero from the negative side when $k\rightarrow 0$ \cite{jungnickel96,eichhorn15}. 
Simple argument shows that this is an artefact of the power series
truncation (for a recent analysis in O(N) models see
Ref.\refcite{Mati:2014xma}). Actually, resummation of any power series expansion (for instance with help of Pad\'e approximants \cite{jakovac15}) should reconstruct the correct asymptotic behaviour of $U_r$ at the
fixed point, which is $\sim \rho_r^{\frac d{d-2}}$, as it can be seen from
asymptotic analysis of (\ref{Z-RGE-B1}). Therefore $U_r =
f(\rho_r) \rho_r^{\frac d{d-2}}$, where $f(\rho_r)\to1$ as
$\rho_r\to\infty$. As a consequence, the coefficients of the power
series expansion of $f(\rho_r)$ have alternating sign. This artefact appears explicitly when in section 5.3 the term $\lambda_{3k}\rho^3/3$ is included into $U_r$. The infrared values of the physical quantities are not affected by the seemingly unstable behaviour of the odd truncations.

The effect of a different class of irrelevant operators on the stability bound can be studied assuming a non-trivial $\rho$-dependence of the Yukawa coupling. In the above explicitly described simplest non-trivial example of involving an irrelevant operator one might test its effect on the stability bound by starting the RG-flow with  $ h_0(k=\Lambda)\approx 0$ and $h_1(k=\Lambda) > 0$ and tuning $h_0(k=0)+h_1(k=0)\rho_{0,min}$ to the value determined by the fermion mass.

\section{Main result: non-perturbative extension of the stability range through an irrelevant Yukawa-type coupling}

\begin{figure}
\begin{center} 
\includegraphics[width=7.0cm]{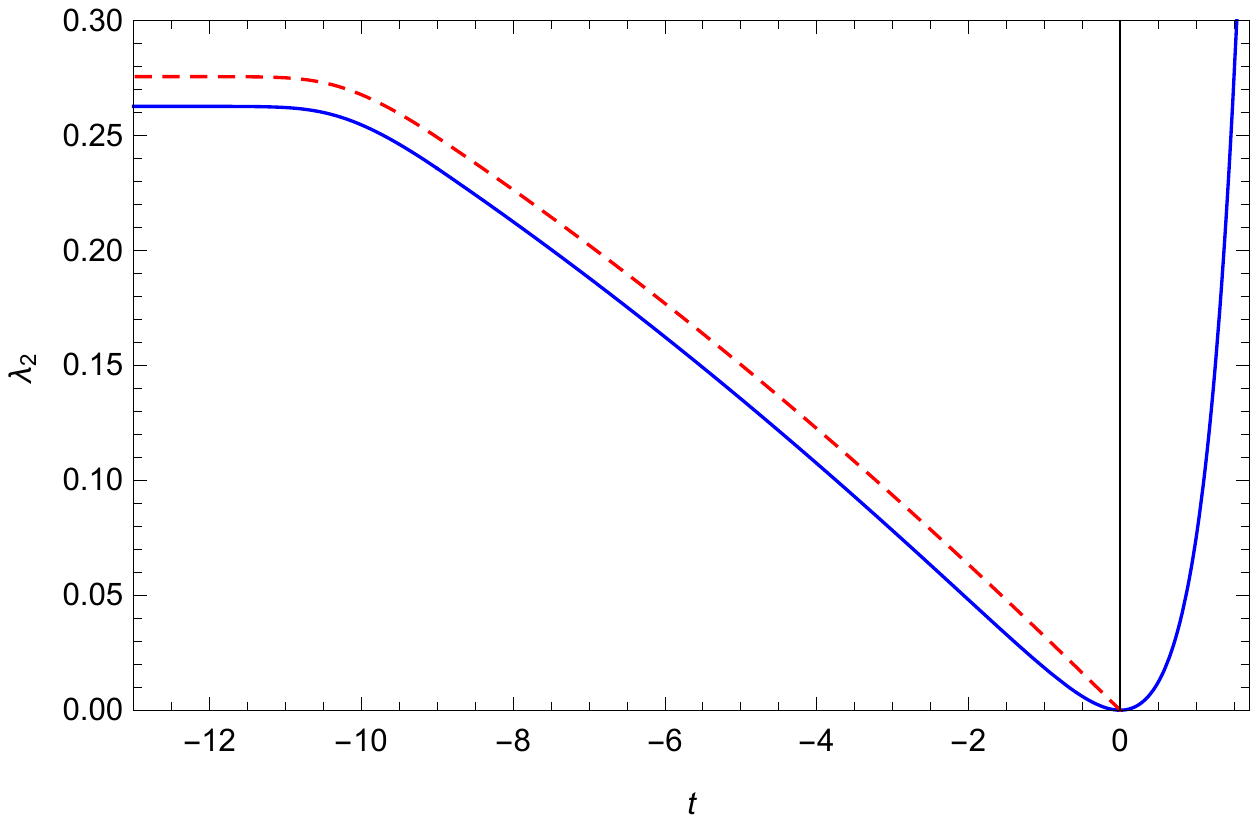}
\end{center}
\caption{The evolution of $\lambda_2$ towards lower scales ($t<0$) for $\Lambda=10^7$ GeV starting with $\lambda_2(k=\Lambda)=0$. The dashed line corresponds to $h_1\equiv 0$. The continuous line represents the evolution for $h_1(\Lambda)=h_{1,max}$ when $\lambda_2$ starts with zero slope. In this case one can smoothly evolve the couplings into the region $t>0$.}
\label{lambda-2-evolution-new}
\end{figure}

\begin{figure}
\begin{center} 
\includegraphics[width=11.5cm]{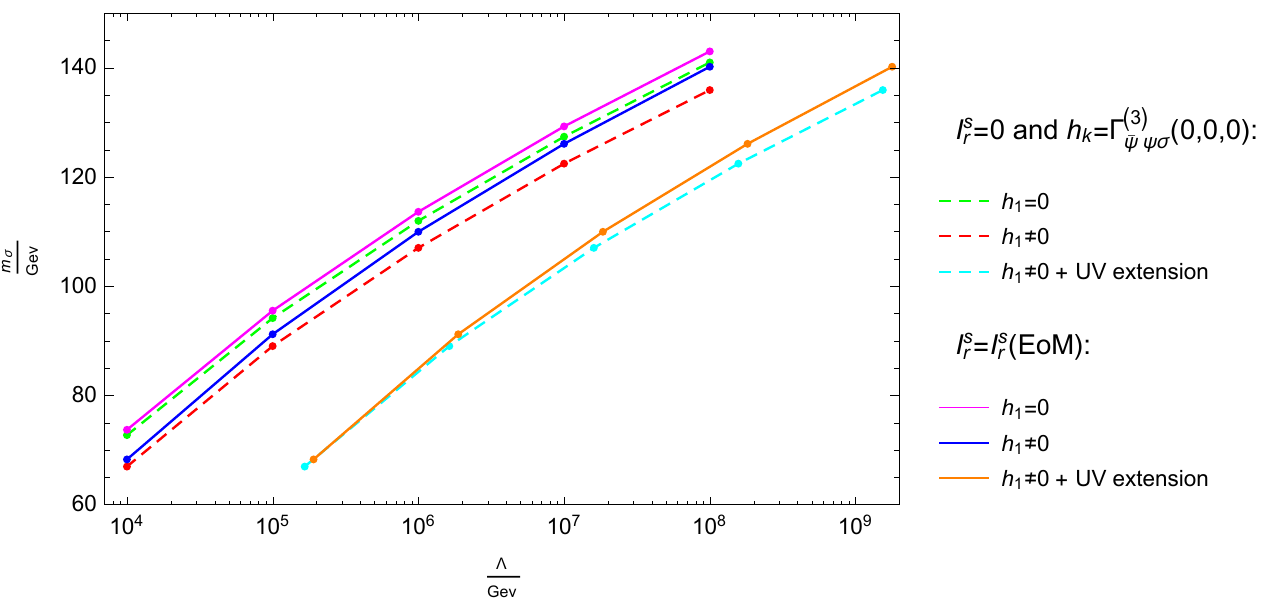}
\end{center}
\caption{The upper left lying two continuous curves represent the lower bounds 
$m_{\sigma, min}(\Lambda)$ obtained for $h_1=0$ and for $h_1=h_{1,max}$, starting the RG-evolution from the scale $k=\Lambda$ and using the expansion point $I_r^s(EoM)$. The considerably lower bound curve in the middle of the figure is obtained by extending the allowed momentum range dynamically to $\Lambda'=\Lambda\exp(t_{max})$. Also the three corresponding bounds arising with the RGE of $h(\rho)$ derived by Gies {\it et al.} and using $I_r^s=0$ are presented (dashed curves).}
\label{lower-mass-bound-new}
\end{figure}

First, we summarize the findings of Ref.\refcite{gies14} concerning the edge of the stability range  without perturbatively irrelevant operators.
In agreement with the one-loop perturbative analysis the lower bound from the RGE's of the system stems from a set of initial data characterised by $\lambda_2(\Lambda)=0$. The quartic coupling evolves from here essentially logarithmically (see the dashed line of Fig.\ref{lambda-2-evolution-new}) until it is saturated with its infrared ($k=0$) value at around $-t\geq 10$, which determines the physical scalar mass. The resulting curves of the cut-off dependence of the lower scalar mass bound computed with one of the two values of $I_r^s$ differ from the results of Ref.\refcite{gies14} only slightly as one can see looking at the two nearby lying curves corresponding to $h(\rho)=h_0$ in the left hand side of Fig.\ref{lower-mass-bound-new}. The difference is the consequence of the different definition of $h_k$, mentioned above and leading to different RGE's beyond the leading one-loop beta-function.  The choice of the expansion point $I_r^s$ has even less (fully negligible) effect: the respective curves coincide within .1\%, therefore cannot be resolved in the figure.

Concerning the influence of including perturbatively irrelevant operators we do not discuss in this paper standalone the effect of switching on higher powers of $\rho_r$ in $U_r(\rho_r)$ which was investigated earlier in Refs.\refcite{gies14,chu15}. We concentrate on allowing linear $\rho$-dependence in the Yukawa-coupling ($h_1\neq 0$) which by itself leads to interesting and potentially important consequences.

The non-trivial running of $h_1$ has the effect of diminishing the starting slope of the $\lambda_2$-evolution, when one starts from $\lambda_2(\Lambda)=0$ towards the infrared (negative $t$-values). The starting slope gradually decreases with increasing $h_1(\Lambda)$ (see the continuous line in Fig.\ref{lambda-2-evolution-new}). It rectifies to its original value with a slight delay, reducing the value of the lower bound belonging to the chosen cutoff by a small amount. The largest acceptable 
$h_1(\Lambda)$ value (denoted as $h_{1,max}$) corresponds to $\lambda_2$ starting horizontally, since beyond this value $\lambda_2(k)$ would (temporarily) cross into the unstable regime. The corresponding slightly lower bounds lie below, but close to the bounds obtained with $h_r(\rho_r)=h_0$.

\begin{figure}
\begin{center} 
\includegraphics[width=7.0cm]{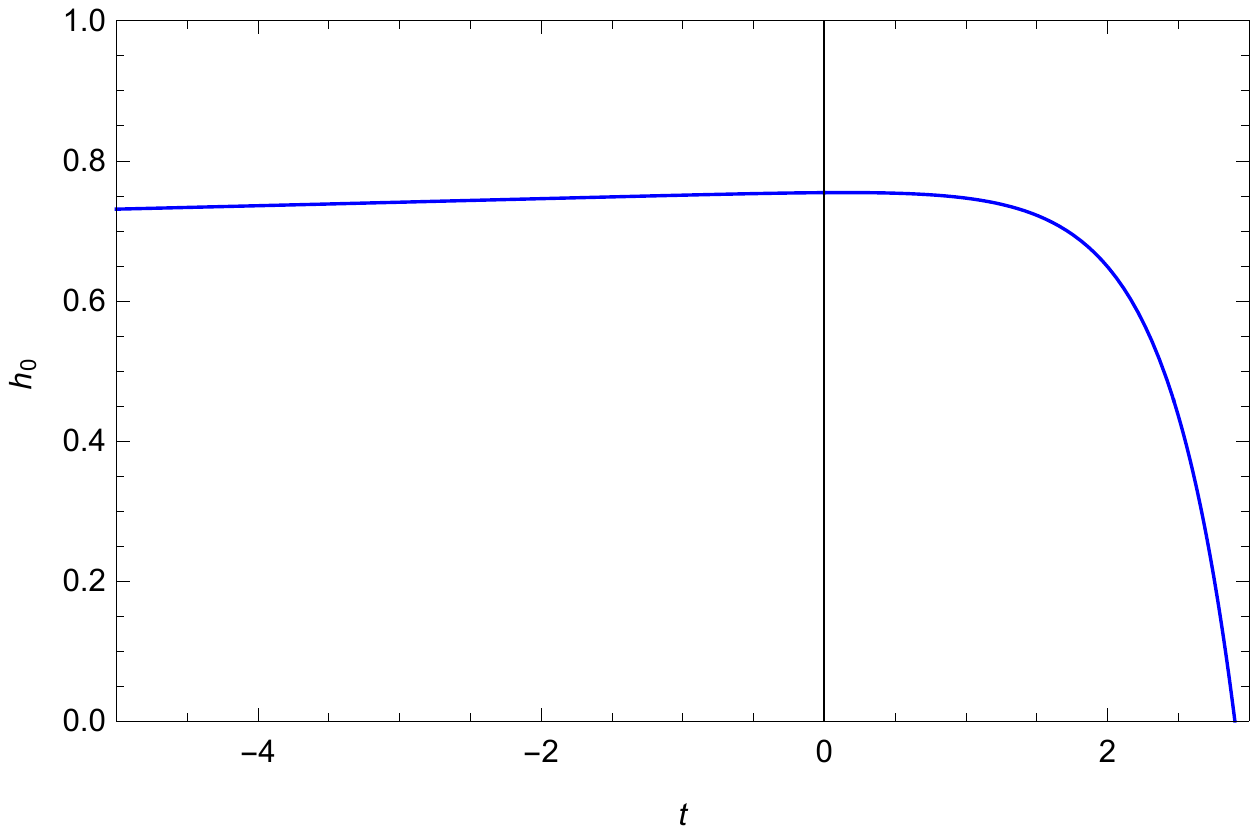}
\end{center}
\caption{The zero of $h_0$ arising when one evolves it into the $t>0$ region stops the dynamical extension of the cutoff theory}
\label{h-0-evolution-new}
\end{figure}

For an essential decrease in the lower bound the following observation is of central importance. In the zero slope case one can continuously integrate the RGE's also upwards ($t>0$). One finds a steeply increasing curve for $\lambda_2(t>0)$ also displayed in Fig.\ref{lambda-2-evolution-new}. This segment of the evolution 
of $\lambda_2$ is controlled by the competition of two terms on the right hand side of its RGE. The term $\sim -h_ 0^4$ comes from the fermion quadrangle and tends to diminish $\lambda_2$. The other term $\sim +h_0h_1$ increases $\lambda_2$ and arises from projecting the contribution of the fermion bubble on the $\sim \rho^2$ term of $U_r$. It turns out that $h_0$ first stays constant, then for $t\geq 1$ steeply decreases (see Fig.\ref{h-0-evolution-new}). On the other hand $h_1$ displays $\sim k^2$ scaling behavior, therefore dominates for $t>0$. As a consequence $\lambda_2$ is expected to scale $\sim k^2$, which in fact is observed. Such a change of the scaling regime of $\lambda_2$ when turning from $t<0$ to $t>0$ is not exceptional. A qualitatively similar change also happens around the point $\lambda_2=\beta_{\lambda_2}=0$, "forced" on the evolution of $\lambda_2$ by coupling the Standard Model to an asymptotically safe quantum gravity theory\cite{shaposhnikov10}.

The evolution for $t>0$ eventually results in a strongly coupled theory with nonzero scalar condensate. The value of the condensate is ${\cal O}(\Lambda'=\Lambda\exp(t))$ and with it both the scalar and the fermion masses are of the order of $\Lambda'(t)$: the scale hierarchy disappears towards the ultraviolet. The integration is stopped at the effective cutoff $\Lambda'_{max}(\Lambda)$, when the steeply decreasing value of $h_0$ (see Fig.\ref{h-0-evolution-new}) reaches zero (a negative Yukawa-coupling (for any value of $\rho$) would destabilize the theory). In our experience no Landau-type (triviality) singularity showed up in $\lambda_2$ until this point was reached.

With the above procedure one determines dynamically an effective cut-off value $\Lambda'_{max}$ with the corresponding couplings (among them $h_0(\Lambda'_{max})=0$) where one actually can start the RG-evolution. One arrives at the same infrared data as if one would start at $t=0$, but now the low-scale (physical) data are hierarchically separated from the initial data set. The new piece of the RG-trajectory, which one can call as strongly coupled "prehistory", causes a substantial delay in the start of the logarithmic variation of $\lambda_2$, and results in a significant extension of the physical momentum range (see the right hand curve in Fig.\ref{lower-mass-bound-new}). The RGE-variant of Ref.\refcite{gies14} can also be treated this way and the emerging new lower mass bound (which appears first in the present paper) is also displayed in Fig.\ref{lower-mass-bound-new}.

Before closing this section we extend our analysis to the $\lambda_2(\Lambda)>0$ case. In the light of the above discussion it is obvious that there exists also in this case a specific $h_{1tuned}$ choice for which $\beta_{\lambda_2}(\Lambda)=0$. More than this is true: a whole interval $h_{1min}<h_{1tuned}<h_{1max}$ can be stretched where the "partial" fixed point and the change in the scaling of $\lambda_2$ occurs, just shifted to slightly higher or lower scales ($k_{FP}$). The corresponding value of $\lambda_2(k_{FP})$ is below $\lambda_2(\Lambda)$, but is required to stay non-negative. This observation makes it clear, that a possibility for smooth UV-extension works for the full allowed infrared range of the Higgs-masses without any need to fine-tune $h_1$. The need for some sort of fine-tuning appears only when one aims at finding the lower mass-bound.

\section{Robustness of the UV-extension within a wider operator set}

In the previous section it was demonstrated that the effect of $h_1$ on the evolution of $\lambda_2$ leads to the emergence of a "partial" fixed point where $\beta_{\lambda_2}=0$. Starting from this point a smooth continuous RG-extension of the validity range of the theory is possible towards the ultraviolet. In this section we present a number of evidences for the robust nature of the appearence of the "partial" fixed point of $\lambda_2$ also upon taking into account the anomalous dimensions of the fields (LPA') and including further perturbatively irrelevant operators into the RGE analysis.
 
\subsection{The impact of anomalous dimensions}

\begin{figure}
\begin{center}
\includegraphics[width=9.5cm]{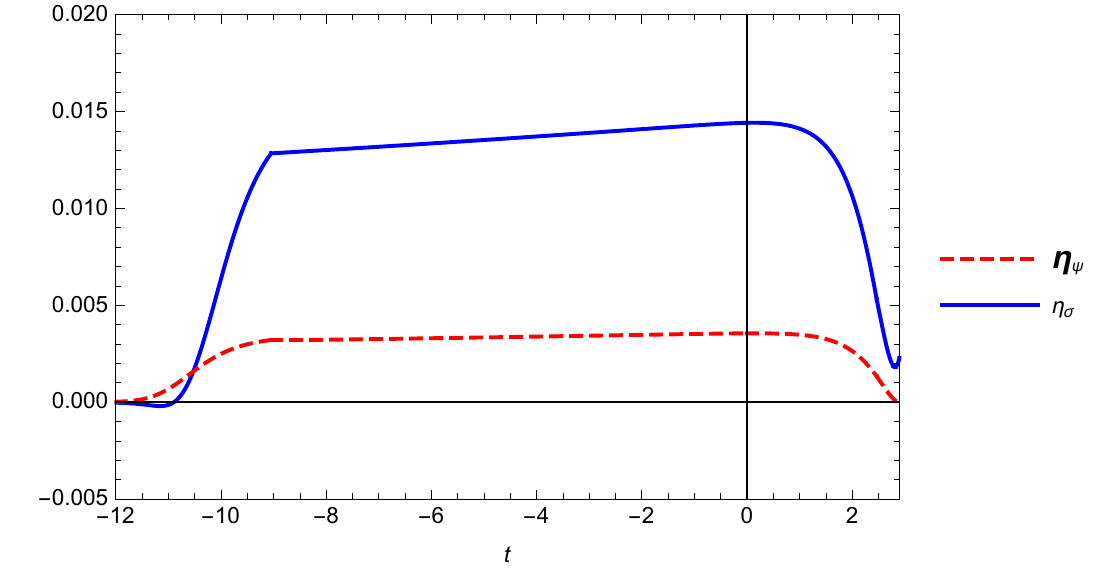}
\end{center}
\caption{The RG-evolution of the anomalous dimensions $\eta_\sigma$ and $\eta_\psi$  ($\lambda_2(\Lambda)=0, \Lambda=10^7$GeV. Initial values of other couplings are tuned as indicated in the main text.)}
\label{eta-t}
\end{figure}

The two RGE's characterising the evolution of the effective potential are completed by the algebraic equations which determine $\eta_\psi, \eta_\sigma$.
The projection of (\ref{super-tracelog-1}) on the coefficients of the kinetic terms by consistency requires the evaluation of the right hand side at some specific value of the background fields. Two expansion points are tested in this paper. Either one chooses $I_k^s\equiv 0$ or we use the point solving the field equation at scale $k$: 
\begin{equation}
\frac{\delta\Gamma_k}{\delta\sigma}\bigl|_{\sigma_k,\psi_k}=h^{(1)}_k(\bar\psi\psi)_k+\sigma_kU_k'(\rho_k)=0\rightarrow
h_k^{(1)}I_k^s(EoM)+2\rho_kU'_k(\rho_k)=0,
\label{effective-field-equation}
\end{equation}
both imply $I^s_{k,min}=0$. This choice leads back to the RG-equations for $\eta_\psi$ and $\eta_\sigma$ determined by \cite{gies14}:
\be
\eta_\psi=2 v_d h_r^2\frac{1}{(1+\tilde\mu_\sigma^2)^2(1+\mu_\psi^2)}\left(1-\frac{\eta_\sigma}{d+1}\right)\Biggl|_{\rho_r=\rho_{r,min}},
\label{eta-psi}
\ee 
\bea
&&
\eta_\sigma=2v_d\Biggl(
\left[
\frac{\left(3U_r''+2\rho_rU_r'''\right)^2}{(1+\tilde\mu_\sigma^2)^4}
-\frac{8h_r^4}{(1+\mu_\psi^2)^4}
\right]\rho_r
\nonumber\\
&&~~~~~~
-\frac{2 h_r^2}{(1+\mu_\psi^2)^2}\left(1-\frac{2}{1+\mu_\psi^2}\right)\left[\frac{1}{1+\mu_\psi^2}+\frac{1}{2}+\frac{1-\eta_\psi}{d-2}\right]\Biggr)\Biggl|_{\rho_r=\rho_{r,min}}.
\label{eta-sigma}
\eea
Here $\tilde\mu_\sigma^2$ is the scaled scalar mass with no correction from $I_r^s(\rho_r)$!
Since the RG-evolution of the model ends in the broken symmetry phase where both fields are massive and therefore $\mu_\psi^2, \tilde\mu_\sigma^2\sim k^{-2}$, these equations imply that $\eta_\psi, \eta_\sigma\rightarrow 0$ when $k\rightarrow 0$. The anomalous dimensions actually might influence the evolution of $U_r$  and $h_r$ only at intermediate scales. The $\eta$-evolution is exemplified in Fig.\ref{eta-t}  where for $\Lambda=10^7$GeV  and $\lambda_2(\Lambda)=0$ the scaled (dimensionless) values of the other three couplings are tuned to the values $\lambda_1(\Lambda)=0.007188, h_0(\Lambda)=0.7546, h_1(\Lambda)=0.4295$, in order to have the physical values of $m_\psi$ and $v$ in addition to arranging $\beta_{\lambda_2}(\Lambda)=0$.

From (\ref{eta-psi}) and (\ref{eta-sigma}) one can express the anomalous dimensions algebraically and substituting them into Eqs. (\ref{Z-RGE-B1}) and (\ref{Z-RGE-B2}) one proceeds to the solution of the resulting equations with the same algorithm as applied when $Z_\sigma=Z_\psi=1$ (this is the ${\textrm{LPA}}^\prime$ variant of the local potential approximation (LPA)). Our general experience is that the anomalous dimensions stay very small (see Fig.\ref{eta-t}), therefore the lower mass-bounds obtained in LPA remain unchanged even quantitatively.    

\subsection{The impact of the strength of the operator $\rho^3$}       

\begin{figure}
\begin{center}
\includegraphics[width=10.5cm]{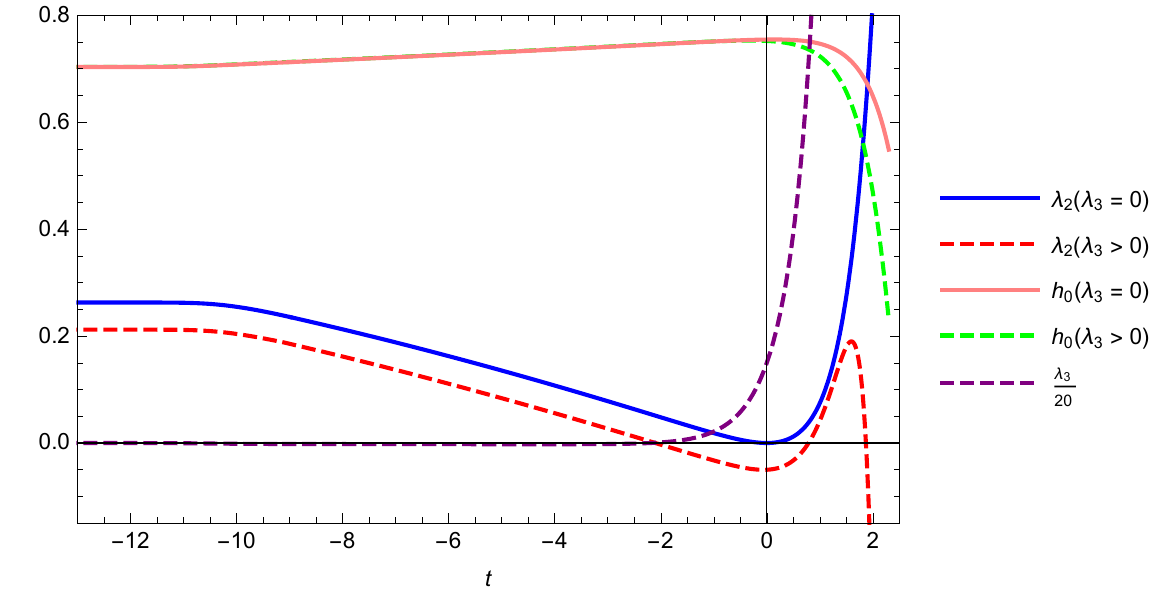}
\end{center}
\caption{Illustration of the effect of switching on $\lambda_3$ on the RG-evolution of $\lambda_2$ and $h_0$ ($\lambda_2(\Lambda)=-0.05, \lambda_3(\Lambda)=3, \Lambda=10^7$GeV. Initial values of other couplings are indicated in the main text).}
\label{lambda_2_lambda3}
\end{figure}

This operator takes over the 
stability control of the potential $U_r$ along the $\rho$-direction from the $\sim\rho^2$ term and lowers considerably the lower Higgs-mass bound as was demonstrated in Refs.\refcite{gies14,chu15}. Here we consider a similar situtation where the RG-evolution towards the infrared starts with $\lambda_2(\Lambda)<0, \lambda_3(\Lambda)>0, \Lambda=10^7$GeV. As one sees in Fig.\ref{lambda_2_lambda3} the freedom to tune $\lambda_1(\Lambda)=0.00736, h_0(\Lambda)=0.756, h_1(\Lambda)=1.1$ allows, in addition to fixing the physical values of $v$ and $m_\psi$, also in this case to tune $\beta_{\lambda_2}(k=\Lambda)=0$. This in turn opens the way to the UV extension of the theory up to some $\Lambda'>\Lambda$. $\lambda_3$ has a diminishing effect on the rate of $\lambda_2$ and eventually compensates at some positive $t$-value the increasing tendency experienced for $\lambda_3=0$. This compensation is followed by a fast crossing of $\lambda_2$ to negative values as one sees in Fig.\ref{lambda_2_lambda3}.  However, in this case the negative value of $\lambda_2$ does not prompt for any stability consideration, since $\lambda_3>0$. The UV-evolution is stopped by the zero of $h_0$ which happens slightly earlier than for $\lambda_3=0$. One also notes the significant decrease in the infrared value of $\lambda_2(k=0)$ relative to the $\lambda_3(\Lambda)=0,\lambda_2(\Lambda)=0$ case. Therefore the effect pointed out in Ref.\refcite{gies14} combines here with the effect of $\rho$-dependent Yukawa-coupling, to lower further the lower Higgs-bound obtained  by applying the two irrelevant operators separately.   

\subsection{The impact of the strength of the operator $\rho^2I$}

The RGE for the coefficient $h_2$ can be deduced by projecting Eq.(\ref{Z-RGE-B2}) on this operator. One tunes also here the couplings $\lambda_1(\Lambda)=0.007168, h_0(\Lambda)=0.756, h_1(\Lambda)=0.4153$ to ensure $m_\psi, v, \beta_{\lambda_2}(\Lambda)=0$. 

It is important to recognize that $h_2$ does not appear directly in the RGE of $\lambda_2$ (one easily figures out that from the fermion-bubble, the lowest power of $\rho$ where $h_2$ contributes to the RGE is $\rho^3$). On the other hand $h_2$ shows up in front of a tadpole type contribution in the RGE of $h_1$. Towards the infrared for moderate $({\cal O}(1))$ values of $h_2$ there is no observable change in $\lambda_2(k=0)$. 

\begin{figure}
\begin{center}
\includegraphics[width=10.5cm]{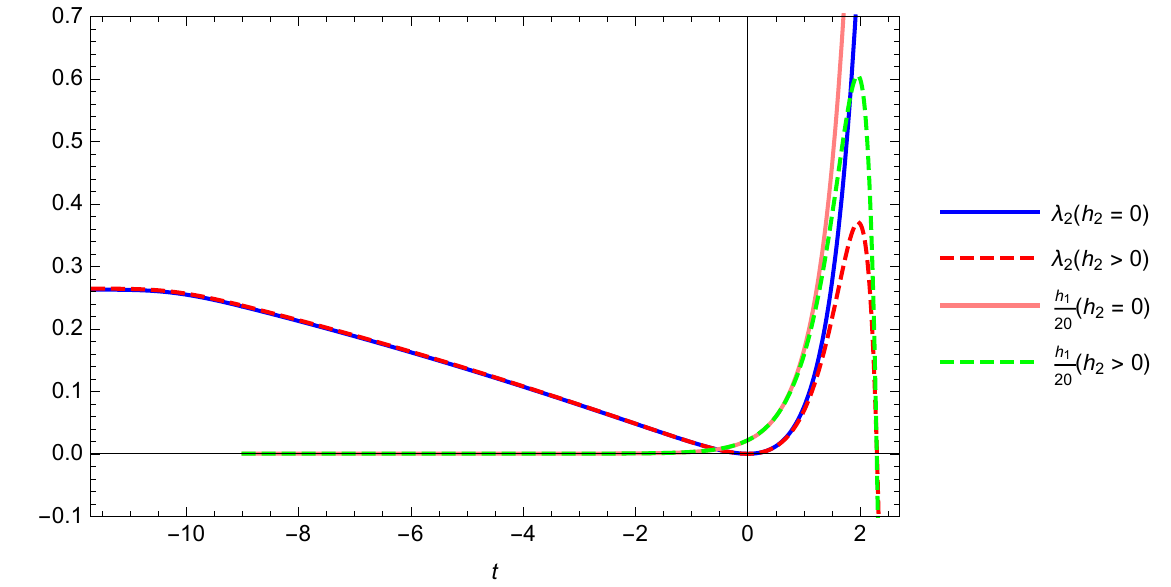}
\end{center}
\caption{Illustration of the effect of switching on $h_2$ on the RG-evolution of $\lambda_2$ and $h_1$ ($\lambda_2(\Lambda)=0, h_2(\Lambda)=0.5, \Lambda=10^7$GeV. Initial values of other couplings are tuned as indicated in the main text.)}
\label{lambda_2_h2}
\end{figure}

The more important question refers to the effect of $h_2$ on the UV ($t>0$) extension of $h_1$. There the increase of $h_1$ is slowed down by its effect, and eventually after reaching a maximum $h_1$ starts to decrease. Since $h_0$ behaves the same way as for $h_2=0$, one reaches a $t$-value where the term $\sim -h_0^4$ compensates $\sim +h_0h_1$. In this point $\lambda_2$ reaches its maximum followed by a fast drop. The  UV-extension is stopped by the zero of $\lambda_2$ or that of $h_0$, occuring first.  This evolution of $\lambda_2$ and of $h_1$ is illustrated in Fig.\ref{lambda_2_h2} by comparing the RG-evolutions with $h_2=0$ and $h_2\neq 0$. Increasing  $h_2(\Lambda)$ pushes the zero of $\lambda_2$ closer to the location of the other zero, the origin (for the present set of parameters the inflection happens for $h_2\approx 65$). Anyhow, there is a natural (and rather wide) range of $h_2(\Lambda)\sim {\cal O}(1)$, where the shift of the curve of the lower mass-bound is the same as for $h_2=0$. 

 Based upon these three specific cases, we conjecture the generic nature of the effect of the operator $h_1\rho I$ on the evolution of $\lambda_2$ resulting in a "partial" fixed point, where $\beta_{\lambda_2}=0$. Starting from this point one is allowed to continue the RG-evolution of $\lambda_2$ towards the ultraviolet. The Landau-singularity of $\lambda_2$ is apparently avoided, the range of extension is limited by the instability caused by other operators.

\section{Conclusions}
The motivation for the presented investigation was to explore the RG-evolution of the two-variable effective potential $W(\rho,I)$ depending also on the fermionic background through $I$. We have derived RGE's for its expansion in $I$ to linear order, and investigated the effect of the choice of the expansion point $I_r^s$.

In this framework it is natural to investigate the case of a genuine $\rho$-dependent coefficient in front of $I$ in the quantum action. The emergence of the perturbatively irrelevant operators is quite natural in the FRG-paradigm. In the absence of a self-consistent UV-completion (governed by a new fixed point), the extended effective action simply describes  a cut-off theory.

We have performed a search for the lower scalar mass bound by tuning the dimension-6 coupling $h_1$, and optimised the extension of the stability range of the theory along this direction. It was demonstrated that the UV-extension of the range of validity of the field theory is made possible through the existence of a "partial" fixed point ($\beta_{\lambda_2}(k_{FP})=0$) for $\lambda_2(\Lambda)>0$ showing up for some scale $k_{FP}$ by choosing the value of $h_1$ from a finite interval. These findings were independently confirmed by the Jena-group\cite{sondenheimer16}. In an extended investigation we also have checked that the location of this interval is slightly shifted when other perturbatively irrelevant operators are also switched on or the anomalous dimensions of the scalar and fermion fields are taken into account. 

The theory arrived at by the UV-extension, at high energy is a strongly coupled theory with a ground state of broken discrete chiral symmetry. All masses are of the order of the cut-off, there is no mass hirarchy in the ultraviolet. The RG-evolution first leads to vanishing $\beta$-function of the quartic scalar coupling $\lambda_2$
at an intermediate scale, wherefrom the theory closely follows the perturbative evolution to the final infrared mass values hierarchically separated from the cutoff. This construction  is hopefully applicable also in other models involving Yukawa-type interaction.

All investigations of the quantum stability of the present model performed to this date point to the possibility of a non-perturbative extension of the allowed momentum range with the application of perturbatively irrelevant operators. One way to further check the persistence of this feature is to continue the gradual expansion of the coupling space along the lines presented in section 5. A more comprehensive approach is to attempt the complete determination of the scale dependent two-variable potential $W(\rho, I)$ with the correct large $\rho$ and $I$ asymptotics as has been realised very recently for the one-variable $U(\rho)$ function\cite{borchardt16}.

\section*{Acknowledgments}
This research was supported by a Hungarian Research Fund Grant K104292. Valuable discussions with H. Gies and R. Sondenheimer, and remarks by J. Pawlowski are thankfully acknowledged. Criticism of anonymous Referees proved in the course of the revision rather constructive.

%\section*{References}

\end{document}